\documentclass[12pt,reprint,tightenlines,groupedaddress,superscriptaddress,floatfix,aps]{revtex4}
\usepackage{graphicx}
\usepackage{rotating}
\usepackage{array}
\usepackage{multirow}
\usepackage{epstopdf}
\usepackage{amsmath}
\usepackage{amsfonts}
\usepackage{setspace}
\usepackage{braket}
\usepackage{color}

\usepackage{hyperref}

\DeclareMathOperator{\Tr}{Tr}

\begin{document}
\title{Interpretation of multiple solutions in fully iterative GF2 and GW schemes using local analysis of two-particle density matrices}

\author{Pavel Pokhilko}
\affiliation{Department  of  Chemistry,  University  of  Michigan,  Ann  Arbor,  Michigan  48109,  USA}
\author{Dominika Zgid}
\affiliation{Department  of  Chemistry,  University  of  Michigan,  Ann  Arbor,  Michigan  48109,  USA}
\affiliation{Department of Physics, University of Michigan, Ann Arbor, Michigan 48109, USA }

\renewcommand{\baselinestretch}{1.0}
\begin{abstract}
Due to non-linear structure, iterative Green's function methods can result in multiple different solutions even for simple molecular systems. 
In contrast to the wave-function methods, a detailed and careful analysis of such molecular solutions was not performed before.
In this work, we use two-particle density matrices to investigate local spin and charge correlators that quantify the charge-resonance and covalent characters of these solutions.
When applied within unrestricted orbital set, 
spin correlators elucidate the broken symmetry of the solutions, 
containing necessary information for building effective magnetic Hamiltonians. 
Based on GW and GF2 calculations of simple molecules and transition metal complexes, we
construct Heisenberg Hamiltonians, four-spin-four-center corrections, as well as biquadratic spin--spin interactions.  
These Hamiltonian parametrizations are compared 
to prior wave-function calculations.
\end{abstract}
\maketitle

Iterative or self-consistent propagator methods that solve the Dyson equation 
are used across diverse areas such as 
nuclear and particle physics\cite{Roberts:Dyson:hadronic:1994,Roberts:Dyson:QCD:2000,Koide:nuclear_dyson:1994}, 
electronic structure of molecules\cite{Stan06,Dahlen05,Ortiz:propagator:review:2013}, 
and condensed matter physics\cite{Mahan00,Negele:Orland:book:2018,Martin:Interacting_electrons:2016}. 
The self-consistent solution of the Dyson equation can be interpreted in the following way. 
A propagating particle interacts with the environment, 
changing it. 
The perturbed environment changes the effective interaction, affecting the particle propagator. 
Therefore, there is a feedback loop between the particle propagation and the environment 
that this self-consistent equation encodes.

The nonlinear equations appearing in iterative Green's function methods may lead to multiple solutions~\cite{Kozik15,Reining:Dyson:multiple_solutions:2017,Rossi:multivalue:2015,Berger:unphys_solutions:2015} fulfilling these equations.
Some of these solutions are discarded as ``unphysical'' and only the remaining  ``physical'' ones should be used to describe systems under study. 
Existence of multiple solutions is a fundamental problem, going beyond convergence difficulties or trapping in local minima. 
The space of acceptable physical solutions is large and deserves a careful analysis. 
In solids, the interpretations of such results is natural since they correspond to different thermodynamic phases. 
In molecular problems, such interpretation is more challenging because the association between these solutions and molecular excited states is not always straightforward when considering only the one-body density matrix (or natural orbital occupations). 
In the wave-function methods, the analysis of many-body wave functions can be done not only via one-electron orbitals\cite{Salem:72:SOGeomDep,ElSayedRule,Michl:ACR:13,Ohrn:73:PROP,Ortiz:99:PROP,Oana:Dyson:07,Luzanov:TDM-1:76,Luzanov:TDM-2:79,HeadGordon:att_det:95,Martin:NTO:03,Luzanov:DMRev:12,Dreuw:ESSAImpl:14,Dreuw:ESSAImpl-2:14,Nanda:NTO:17,Wojtek:ImagEx:18,Krylov:Libwfa:18,Dreuw:NTOfeature:2019,Pavel:SOCNTOs:2019,Krylov:Orbitals,Nanda:RIXSNTO:20,Nanda:NTO:hyperpolarizability:2021}, 
but also by examining the determinantal wave-function structure\cite{Scholes:EET:1994,Toru:ASDDimers:13,Toru:ASD:14} 
or high-order density matrices\cite{Gill:intracule:2003,Gill:intracule:2004,Luzanov:HP:06,Mazziotti:cumulant:norm:06,Plasser:NTOfeature:2020}. 
However, for one-body iterative Green's functions such analysis was limited and could not be done to a comparable level since it requires information beyond the one-particle picture.

In this paper, we analyze GW and GF2 
self-consistent solutions for a few strongly correlated systems by means of local spin and charge correlators. 
This approach, based on the recently derived two-particle density matrices\cite{Pokhilko:tpdm:2021}, 
helps us to rationalize the underlying electronic structure and explain the character of the solutions obtained. 
In particular, it elucidates the broken-spin character of the solutions when applied to magnetic systems, 
establishing the interpretation ground of such calculations. 
Moreover, local spin and charge correlators serve as diagnostic tools for non-Heisenberg type interactions in magnetic species.

An imaginary-time Green's function $G$ is defined as 
\begin{gather}
G_{pq} (\tau) = -\frac{1}{Z} \Tr \left[e^{-(\beta-\tau)(H-\mu N)} p e^{-\tau(H-\mu N)} q^\dagger  \right], \\
Z = \Tr \left[ e^{-\beta(\hat{H}-\mu \hat{N})} \right], 
\end{gather}
where $\tau$ is the imaginary time, $\beta$ is the inverse temperature, $\mu$ is the chemical potential, 
$\hat{N}$ is the particle-number operator, and $Z$ is the grand-canonical partition function. 
Since the fermionic imaginary-time one-particle Green's function is antiperiodic 
and the bosonic Green's function is periodic,  
the corresponding frequency representations are discrete, known as Matsubara frequencies:
\begin{gather}
\omega_n = \frac{(2n+1) \pi}{\beta}, \\
\Omega_n = \frac{2n \pi}{\beta},
\end{gather}
where $\omega_n$ are fermionic Matsubara frequencies and $\Omega_n$ are bosonic Matsubara frequencies. 
The Dyson equation is written as 
\begin{gather}
G^{-1}(i\omega_n) = G^{-1}_0(i\omega_n) - \Sigma[G](i\omega_n),
\protect\label{eq:Dyson}
\end{gather}
where the self-energy $\Sigma[G]$ is a functional of the one-particle Green's function $G$ 
and $G_0$ is the Green's function of independent electrons. 
The exact definition of the self-energy depends on the actual approximation employed.

GF2 is a perturbative method that includes contributions to the self-energy up to the second order in terms of bare Coulomb interaction\cite{Snijders:GF2:1990,Dahlen05,Phillips14,Rusakov16, Welden16, Iskakov19}.  
GW~\cite{Hedin65,G0W0_Pickett84,G0W0_Hybertsen86,GW_Aryasetiawan98,Stan06,Koval14,scGW_Andrey09} approximates the static part of self-energy with Hartree--Fock exchange. 
The dynamic part of its self-energy is defined in terms of screened frequency dependent Coulomb interaction $W(\Omega_n)$.

The self-consistently solved equations lead to the following structure of the two-particle density matrix
\begin{gather}
\Gamma_{\braket{p q |r s}} = 
\gamma_{pr} \gamma_{qs} - \gamma_{ps} \gamma_{qr} + \Gamma^\text{conn}_{\braket{p q |r s}}, \\
\Gamma^\text{disc}_{\braket{p q |r s}} = \gamma_{pr} \gamma_{qs} - \gamma_{ps} \gamma_{qr}, 
\protect\label{eq:Gamma}
\end{gather}
where $\Gamma$ is the two-particle density matrix, 
$\gamma$ is the one-particle density matrix, and 
$\Gamma^\text{conn}$ and $\Gamma^\text{disc}$ are the connected and disconnected parts of two-particle density matrix, 
respectively.  
The connected part is also known as the electronic cumulant of the two-particle density matrix.  

The self-energy functional $\Sigma[G]$, particular for the approximation employed,  determines the actual expression for the electronic cumulant. 
The GF2 and GW approximations lead to the following cumulants:
\begin{gather}
\Gamma_{\braket{p_0 q_0 | r_0 s_0}}^\text{GF2} = 
\frac{1}{\beta}\sum_{\omega_n} 
\sum_{t}  
I^{dir+ex,1}_{p_0 q_0 t s_0}(i\omega_n) G_{tr_0}(i\omega_n),
\protect\label{eq:GF2_cum_antisym} \\
I^{dir+ex,1}_{p_0 q_0 t s_0}(\tau) = 
- \sum_{uvw} \braket{tu || vw} G_{vp_0}(\tau) G_{wq_0} (\tau) G_{s_0u} (-\tau) \protect\label{eq:I1}, \\
\Gamma_{(p_0 q_0 | r_0 s_0)}^\text{GW} = 
-\frac{1}{\beta}\sum_{\Omega_m} 
\sum_{pqrs} \Pi_{r_0 s_0 pq}(\Omega_m) W_{(pq|rs)}(\Omega_m) \Pi_{rs p_0 q_0}(\Omega_m), 
\end{gather}
where all the indices label spin-orbitals.

\begin{figure}[!h]
  \includegraphics[width=7cm]{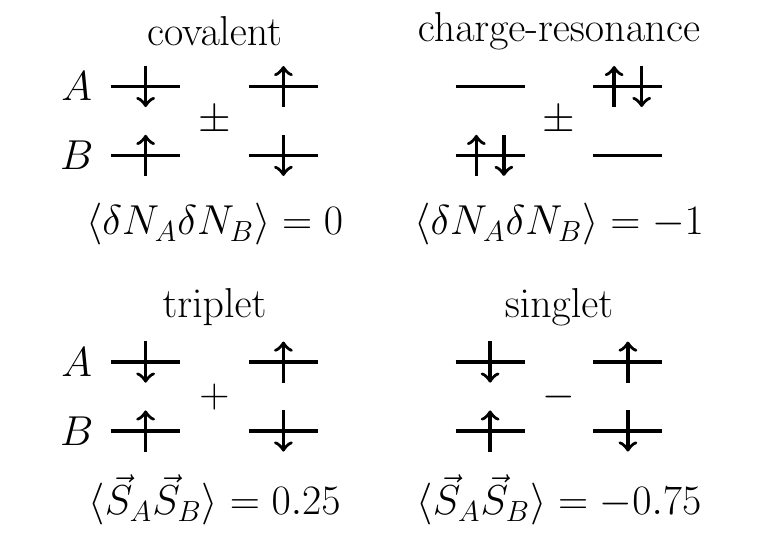}
\centering
\caption{
Possible configurations formed with 2 electrons and 2 localized orbitals $A$ and $B$. 
Local different-center charge and spin correlators are shown.
         \protect\label{fig:config}
}
\end{figure}

A statistical cumulant\cite{Kubo:cumulant:1962} of two operators $\hat{X}$ and $\hat{Y}$ is defined as
\begin{gather}
\Pi_{XY} = \braket{\hat{X}\hat{Y}} - \braket{\hat{X}}\braket{\hat{Y}} = 
\braket{(\hat{X} - \braket{\hat{X}}) (\hat{Y} - \braket{\hat{Y}})} =
\braket{\delta\hat{X} \delta\hat{Y}}.
\end{gather}
Statistical cumulants are also known as covariances. 
They measure the joint variability of two random variables.  
Covariances employ the mean value of the product of the
deviations of two variables from their respective means/averages 
(here denoted as $\braket{\hat{X}}$ and $\braket{\hat{Y}}$). 
When deployed within localized orbitals, 
statistical cumulants of local operators give a local into the probe of electronic structure. 
For example, local charge cumulants have been used to define 
a chemical bond order\cite{Ruedenberg:chem_bond:1962,Jorge:bond_index:1985,Torre:popul:cumulants:2002,Bochicchio:bond_order:2003,Goddard:corr_chem_bond:1998,Luzanov:bond_indices:2005,Mayer:bond_order:2007} 
and local spin cumulants have been used for the analysis of 
polyradical species\cite{Davidson:local_spin:2001,Davidson:local_spin:2002,Davidson:MolMagnets:2002,Hess:local_spin:2005,Luzanov:SpinCorr:15}. 
Here the local operators are constructed from the integrals, expressed in orbitals localized~\cite{note:nonorth} 
in a certain space region. 
We follow the Mulliken-like partitioning, similar to  Ref.\cite{Goddard:corr_chem_bond:1998,Bochicchio:bond_order:2003,Hess:local_spin:2005}, 
i.e. use atomic orbitals for partitioning of the system into fragments. 
Consequently, the local particle number and local spin operators on a fragment $A$ are defined as
\begin{gather}
\hat{N}^A = \sum_{pq \in A} \braket{p|q} ( a^\dagger_{p,\alpha} a_{q,\alpha} + a^\dagger_{p,\beta} a_{q,\beta}) \\
\hat{S}_z^A = \frac{1}{2}\sum_{pq \in A} \braket{p|q} ( a^\dagger_{p,\alpha} a_{q,\alpha} - a^\dagger_{p,\beta} a_{q,\beta}) \\
\hat{S}_+^A = \sum_{pq \in A} \braket{p|q} a^\dagger_{p,\alpha} a_{q,\beta} \\
\hat{S}_-^A = \sum_{pq \in A} \braket{p|q} a^\dagger_{p,\beta} a_{q,\alpha}  
\end{gather}

Below we consider a few examples that can be rationalized by means of local charge and spin correlators. 
Cartesian geometries of each molecule is given in the Supplementary Information. 
Dunning's cc-pVDZ basis sets were used in all calculations\cite{Dunning:ccpvxz:1989,Dunning:ccpvxz:Al-Ar,Duninng:ccpvxz:Sc-Zn}. 
Green's function calculations were performed with the in-house GF2 and GW codes 
within resolution-or-identity (RI) approximation of two-electron integrals\cite{Rusakov16,Iskakov20}; 
integrals were computed with PySCF program\cite{PYSCF}. 
Full configuration interaction (FCI) was performed with GAMESS program (2020 R1 version)\cite{GAMESS}.  

\begin{figure}[!h]
  \includegraphics[width=7cm]{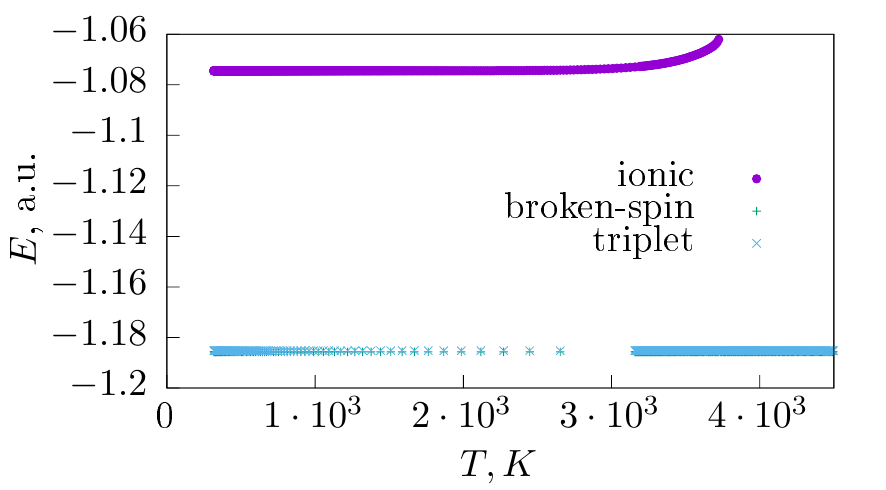}
  \includegraphics[width=7cm]{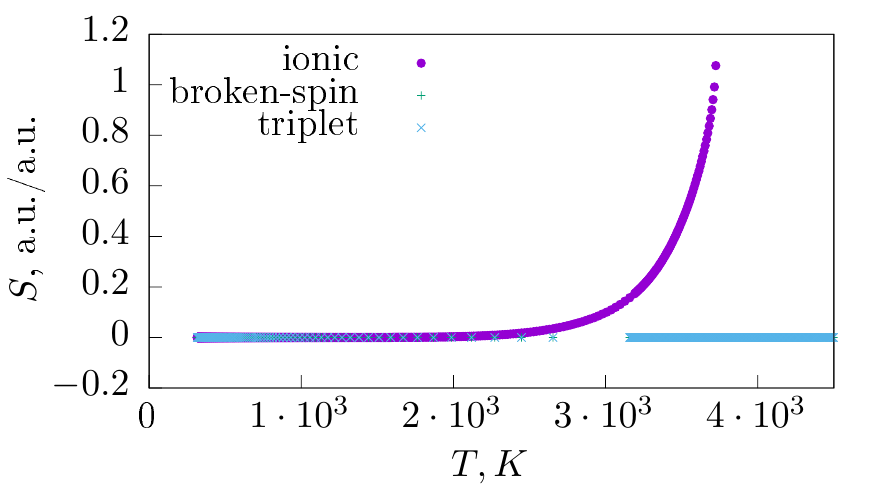}
\centering
\caption{The temperature dependence of the GW internal energy and entropy for the H$_2$ molecule. 
The solutions are labeled according to the physical meaning of the corresponding HF solutions.
         \protect\label{fig:GW_T}
}
\end{figure}
\textbf{H$_2$ molecule.} We obtained 3 zero-temperature Hartree--Fock solutions for a stretched hydrogen molecule: 
broken-spin UHF, high-spin UHF (triplet), and RHF (ionic) solutions. 
Using these HF solutions as guesses, we converged the corresponding unrestricted and restricted 
finite-temperature self-consistent HF, GW, and GF2 calculations.  
The resulting finite-temperature solutions are distinct from each other. 
To obtain these solutions as a function of temperature, the chemical potential is optimized at $\beta=1000$~a.u.$^{-1}$ separately for each of the solutions to ensure that $n=2$, where $n$ is the number of electrons. 
Subsequently, the chemical potential was fixed at this value for all other temperatures.  
The energies and entropies for the GW solutions (Figure~\ref{fig:GW_T}) are flat in a broad range of temperatures. 
At a very high temperature, the GW solution obtained from RHF (ionic) solution  
starts losing electrons (due to maintaining a constant chemical potential for all temperatures), which leads to an increase in the energy and entropy (see Figure~\ref{fig:GW_T}). 

\begin{table*} [tbh!]
  \caption{Local charge and spin correlators for the H$_2$ molecule, 
           computed for Green's functions at $\beta=1000$~a.u. 
           Both full GW and only disconnected part of the GW correlators are are listed. }
\protect\label{tbl:H2_corr}
\begin{tabular}{l|ccc|l|ccc}
\hline
\hline
\multicolumn{4}{c|}{$\mathbf{\braket{\delta N_A \delta N_B}}$ }& \multicolumn{4}{c}{$\mathbf{\braket{\delta N_A \delta N_A}}$}\\ 
\hline
\hline
method & ionic   & triplet & broken-spin & 
method  & ionic   & triplet & broken-spin \\
\hline
HF (finite temp)      & -0.38  & 0.00  & 0.00 &  HF  (finite temp)     & 0.50 & 0.00 & 0.00 \\
GW, disc & -0.28  & 0.00  & 0.00 &  GW, disc & 0.51 & 0.01 & 0.01 \\
GW       & 0.13   & 0.00  & 0.00 &  GW       & 0.64 & 0.30 & 0.30 \\
GF2      & -0.17  & 0.00  & 0.00 &  GF2      & 0.21 & 0.00 & 0.00 \\
\hline
\hline
\multicolumn{4}{c|}{$\mathbf{\braket{\bar{S}_A \bar{S}_B}}$}& \multicolumn{4}{c}{ $\mathbf{\braket{\bar{S}_A \bar{S}_A}}$}\\ 
\hline
\hline
method  & ionic   & triplet & broken-spin &
method  & ionic   & triplet & broken-spin \\
\hline
HF (finite temp)      & -0.39  & 0.25  & -0.25 & HF  (finite temp)     & 0.38 & 0.76 & 0.76 \\ 
GW, disc & -0.28  & 0.25  & -0.25 & GW, disc & 0.39 & 0.76 & 0.76 \\ 
GW       & -0.07  & 0.25  & -0.25 & GW       & 0.60 & 0.83 & 0.83 \\
GF2      &  0.00  & 0.25  & -0.25 & GF2      & 0.61 & 0.76 & 0.76 \\
\hline
\end{tabular}
\end{table*}

Tab.~\ref{tbl:H2_corr} shows local charge and spin correlators between the same and different centers. Values of these correlators can be used to interpret and distinguish different molecular solutions as illustrated in Fig.~\ref{fig:config}. 
The zero-temperature RHF (ionic) solution is a mixture of covalent ($\braket{\delta N_A \delta N_B}=0$) and charge-resonance configurations ($\braket{\delta N_A \delta N_B}\ne0$). 
Correlators, computed with the finite-temperature RHF solution, also show mixed charge-resonance and covalent characters. 
Characters of both finite-temperature UHF solutions also match the corresponding zero-temperature UHF solutions. 
These trends can be explained through a symmetry breaking of a particular type:
\begin{gather}
\hat{\rho} = \sum_{i} w_i \ket{\Psi_i}\bra{\Psi_i} \rightarrow 
\hat{\rho}_{HF}^{(T=0)} = \ket{\Phi_{HF}}\bra{\Phi_{HF}}, 
\protect\label{eq:HF_rho}
\end{gather}
where $\hat{\rho}$ is the grand canonical density operator, $w_i$ are the Boltzmann weights for an eigenstate $\Psi_i$, 
and $\Phi_{HF}$ is the Hartree--Fock determinant (the last equality is written in the low-temperature limit). 
In general, when the symmetry breaking occurs, it does not result in a pure-state low-temperature limit, 
as is the case, for example, for HF with fractional occupancies. 
The corresponding symmetry analysis of $\hat{\rho}$ requires consideration of \emph{projective} representations of symmetry groups rather than linear representations.  
This gives more pathways of symmetry breaking, 
going beyond Fukutome's HF symmetry breaking classification\cite{Fukutome:UHF:81} (see also generalization for Green's functions\cite{Mochena:Fukutome:broken_symmetry:GF}).  
However, in our case, we observe symmetry breaking with the pure-state low-temperature limit
which is easy to analyze.  

At zero temperature, the high-spin triplet UHF determinant describes the triplet wave function qualitatively correctly. 
However, the corresponding finite-temperature $\hat{\rho}$ should contain contributions 
from triplet configurations with all spin projections (since their energies are degenerate), 
requiring a multiconfigurational treatment. 
This results in a triplet symmetry-broken finite temperature UHF solution with unequal average numbers of $\alpha$ and $\beta$ electrons while 
having a triplet pure-state UHF solution in the low-temperature limit. 
For the finite temperature symmetry-broken triplet, local spin correlators are $0.76$ and $0.25$ for the spins on the same and different centers, respectively.
The broken-spin UHF determinant leads to a broken-spin finite-temperature UHF solution with equal average numbers of 
$\alpha$ and $\beta$ electrons. 
While the local spin correlator on the same centers is $0.76$, 
the local spin correlator between different centers is $-0.25$ due to the opposite orientation of local spins in the broken-spin solution.   
Local charge correlators for both UHF solutions are zero, confirming the covalent character of these solutions.

GW and GF2 preserve the characters of the broken-spin UHF, high spin UHF, and RHF ionic solutions. 
Their energy and entropy temperature dependences are too flat (see Figure~\ref{fig:GW_T}) because these methods do not provide enough correlation to properly illustrate temperature dependencies of these solutions (FCI does not have this issue as shown in Fig.~S1 in SI). 
Although weakly correlated methods can give expressions for the density operator analogous to Eq.~\ref{eq:HF_rho},
one should be careful with such generalizations, since Green's function methods 
do not have to be ensemble representatable. 
For example, the GW two-particle density matrix does not have the permutational symmetry 
 required to ensure ensemble representability. 
Yet, local correlators provide useful characterizations for the solutions obtained. 
Different-center spin correlators evaluated for GW and GF2
obtained from the RHF (ionic) guess become closer to zero.
The different-center charge correlators for this ionic solution range from $-0.38$ (HF) to $0.13$ (GW) and $-0.17$ (GF2) indicating that the presence of electronic correlation suppresses the high-energy ionic contributions 
and enhances the covalent character of the solution. 
This is also suggested by the increase in the local 
same-center spin correlators from $0.38$ (HF) to $0.60$ (GW) and $0.61$ (GF2). 
Although these correlators show an improvement in the description of strong correlation, 
the deviation from their limiting values is still significant, indicating that the solutions are partly contaminated with the ionic contributions. 
The above analysis explains why the restricted GF2 does not give correct electronic energies at the dissociation limit. 

GF2 and GW solutions obtained from the UHF guesses preserve the character of the 
corresponding Hartree--Fock solutions. 
However, same-center GW correlators (the charge and spin correlators are 0.30 and 0.83, respectively) show deviations from the expected values. 
This artifact appears only when the electronic cumulant is present indicating that it is
 likely due to the omission of the correct permutational structure of the GW cumulant. 
This observation is consistent with our earlier findings of large charge and spin fluctuations in the atomic GW calculations\cite{Pokhilko:tpdm:2021}. 
The GF2 cumulant has a correct permutational structure and does not lead to these artifacts. 
Since the same-center correlators for the triplet and broken-spin GW solutions are numerically close, 
we hope that these effects will be cancelled when energy differences are considered.  

The interpretation of energies of broken-spin UHF and DFT solutions was pioneered by Noodleman\cite{noodleman:BS:81}, 
who suggested to construct effective Heisenberg Hamiltonians from such solutions. 
This approach found its widespread use in the evaluations of magnetic couplings for molecular magnets\cite{Noodleman:BS:1986,Ruiz:BSDFT:99,Nishino:BSDFT:97,Soda:BSDFT:00,sinnecker:BS:Mn:04}.  
However, a rigorous theoretical justification of such a scheme is not always clear since the evaluation of two-particle properties in DFT is usually done for the Kohn--Sham determinant, 
which is not exact\cite{Cremer:DFT:S2:2001,Handy:DFT:S2:2007,Vedene:DFT:S2:1995}.  
Here, we are following the Davidson's and Clark's\cite{Davidson:local_spin:2001,Davidson:local_spin:2002,Davidson:MolMagnets:2002} reasoning and use local spin correlators to construct the Heisenberg Hamiltonian and use the known relations between the effective exchange coupling $J$ and energy differences:
\begin{gather*}
H = -J_{AB} \mathbf{S}_A \cdot \mathbf{S}_B, \\
J = E(S) - E(T) = 2\left( E(BS) - E(T) \right), 
\end{gather*}
where S, T, and BS denote the singlet, triplet, and broken-spin solutions. 
Such construction of effective Hamiltonian is different from the Bloch's effective Hamiltonian theory\cite{Bloch:1958,Cloizeax:1960,Okubo:1954}, 
also used for derivations of the effective magnetic Hamiltonians\cite{Calzado:02,Guihery:2009,Malrieu:2010,Marlieu:MagnetRev:2014,Mayhall:2014:HDVV,Mayhall:1SF:2015,Pokhilko:EffH:2020,Pokhilko:spinchain}. 

Tab.~\ref{tbl:H_params} contains the extracted effective exchange couplings from the broken-spin and FCI solutions. 
\begin{table} [tbh!]
  \caption{Extracted parameters from FCI and broken-spin solutions in cm$^{-1}$.
\protect\label{tbl:H_params}}
\begin{tabular}{cc|cccc}
\hline
\hline
System & Parameter  &  FCI   &   GW  & GF2  & UHF \\
\hline
\hline
H$_2$  & $J$    &  -213  &  -217 & -218 & -212 \\
\hline                                 
H$_4$  & $J$    &  -185  &   -189      & -189      & -183      \\
H$_4$  & $J_{4c}$  &  -3.9  &   -3.8      & -3.8      & -3.4      \\
\hline
\end{tabular}
\end{table}
The Noodleman's interpretation applied to UHF, GW, and GF2 predicts  exchange constants 
differing from the  FCI ones by only a few wavenumbers.

\textbf{Tetrahedral H$_4$.} Stretched tetrahedral H$_4$ is another example of a strongly correlated system. 
Its Hartree--Fock broken-symmetry solutions has been 
classified before\cite{Fukutome:H4:1975,Scuseria:magnetic:density:2018}.  
Here, we consider only the unrestricted solutions computed from axial UHF guesses. 
Similarly to the stretched H$_2$ molecule, all different-center spin correlators are close to $0.25$ and $-0.25$. 
The low-energy physics can be reconstructed through the following effective Hamiltonian:
\begin{gather*}
H = -J \sum_{i<j} \mathbf{S}_i \cdot \mathbf{S}_j 
- J_{4c} \sum_{\braket{ijkl}} (\mathbf{S}_i \cdot \mathbf{S}_j) (\mathbf{S}_k \cdot \mathbf{S}_l),
\end{gather*}
where the first term is the usual Heisenberg term while the second one is the four-spin-four-center interaction 
describing a deviation from the Heisenberg model. 
The summation in the second term runs over a unique partitioning of such indices: 
(12,34), (13,24), (14,23), where each of the numbers enumerates a radical center.  
The eigenvalues of this Hamiltonian are:
\begin{eqnarray}\nonumber
E(S=2) &=& -\frac{3}{2}J -\frac{3}{16}J_{4c} \\ \nonumber
E(S=1) &=& \frac{1}{2}J + \frac{5}{16}J_{4c} \\ \nonumber
E(S=0) &=& \frac{3}{2}J -\frac{15}{16} J_{4c}. \nonumber
\end{eqnarray}
Broken-spin configurations are described by the diagonal of this model in a basis of determinants. 
The corresponding energies of the broken-spin configurations are:
\begin{eqnarray}\nonumber
E(S_z=\pm 2) &=& -\frac{3}{2}J -\frac{3}{16}J_{4c} \\\nonumber
E(S_z=\pm 1) &=& 0 + \frac{3}{16}J_{4c} \\\nonumber
E(S_z=\pm 0) &=& \frac{1}{2}J -\frac{3}{16} J_{4c}.\nonumber
\end{eqnarray}
\begin{table} []
  \caption{Energies of broken-spin solutions in cm$^{-1}$. 
The high-spin solutions are taken as a zero energy.  
\protect\label{tbl:H4_bs_energies}}
\begin{tabular}{c|ccc}
\hline                        
\hline                        

          &   GW        & GF2       & UHF \\
\hline                        
\hline                        

$S_z=0$   &   -378      & -379      & -366      \\
$S_z=1$   &   -285      & -286      & -276      \\
$S_z=2$   &    0        &  0        &  0        \\
\hline
\end{tabular}
\end{table}
There are many degenerate broken-spin solutions, namely
4 solutions with $S_z = 1$, 
6 solutions with $S_z = 0$, 
and only one high-spin solution with $S_z = 2$. 
Such a high degree of degeneracy becomes clear if one looks at linear combinations of the spin-complete wave-functions, 
which mix into broken-spin configurations. 
Each $S_z = 1$ or $S_z = 0$ broken-spin configuration has contributions 
from wave-functions belonging to different irreducible representations of the  $T_d$ tetrahedral point group. 
Thus, the broken-spin configurations form a larger, reducible representation.
The energies of the broken-spin solutions, the extracted $J$ and $J_{4c}$ parameters, 
and the reconstructed energies are shown in Tables~\ref{tbl:H4_bs_energies}, \ref{tbl:H_params}, and \ref{tbl:H4_energies}.
Extracted parameters and interpreted energies of GW, GF2, and UHF are very close to FCI. 
While H$_2$ and H$_4$ are toy examples that provide a basic validation for the exchange coupling constants extracted from UHF, GW, and GF2, 
realistic, many-electron, strongly correlated systems are more complex. 
In these systems, UHF $\braket{S^2}$ may significantly differ from the values, 
expected from the Noodleman's interpretation. 

Below we consider two more complicated transition metal complexes with
electronic structure qualitatively resembling the Heisenberg Hamiltonian. 
\begin{table} [tbh!]
  \caption{FCI energies and constructed energies from broken-spin solutions in cm$^{-1}$.  
The high-spin solutions are taken as a zero energy.  
\protect\label{tbl:H4_energies}}
\begin{tabular}{c|cccc}
\hline    
\hline                                 
Symmetry  &  FCI   &   GW  & GF2 & UHF \\
\hline    
\hline                                 
$^1$E     &  -552  &   -564      & -565      & -546      \\
$^3$T$_2$ &  -372  &   -380      & -381      & -368      \\
$^5$A$_1$ &   0    &    0        &  0        &  0        \\
\hline
\end{tabular}
\end{table}
\begin{table} [tbh!]
  \caption{A comparison between effective exchange constants (in cm$^{-1}$) extracted from the broken-spin GW, 
wave-function theories, and experiments. 
\protect\label{tbl:J_complexes}}
\begin{tabular}{c|ccp{3.1cm}c}
\hline    
\hline                                 
Complex             &  BS-UHF & BS-GW    & WF theories    & Exp.   \\
\hline    
\hline    
Cu$_2$Cl$_6^{2-}$   &  $-57.4$ & $-48.8$ & $-16$ $^a$ & 0, $-40$\cite{Caballol:Cu_struct:1994}   \\
\hline                                 
Fe$_2$OCl$_6^{2-}$  &  $-44.4$ & $-164.1$ & $-125.4$..$-108.6$ $-250.2$..$-226.8$ $-233.6$..$-190.8$ $^b$  & $-234$\cite{Molins:Fe2OCl6:magnet:2003}   \\
\hline
\end{tabular}

$^a$ EOM-SF-CCSD\cite{Orms:magnets:17} \\
$^b$ Ranges of effective exchange constants (CASSCF, MRCI+Q, DMRG, respectively), extracted from the pairs of states with $\Delta S = 1$\cite{Morokuma:DMRG:biquad_exc:2014}. 
\end{table}

\textbf{Transition metal complexes: [Cu$_2$Cl$_6$]$^{2-}$ and [Fe$_2$OCl$_6$]$^{2-}$.} 
Geometries of [Fe$_2$OCl$_6$]$^{2-}$ and [Cu$_2$Cl$_6$]$^{2-}$  were taken 
from their crystal structures~\cite{Molins:fe_struct:1998,Caballol:Cu_struct:1994}, 
and are consistent with previous theoretical studies~\cite{Morokuma:DMRG:biquad_exc:2014,Orms:magnets:17}. 
Similar to the H$_2$ molecule, we obtained the high-spin and broken-spin UHF and GW solutions for both
[Cu$_2$Cl$_6$]$^{2-}$ and [Fe$_2$OCl$_6$]$^{2-}$. 
Using the Noodleman's interpretation of broken-spin solutions, 
the extracted antiferromagnetic effective exchange couplings $J$ 
agree with the experimental results and previous theoretical studies (see Tab.~\ref{tbl:J_complexes}). 
The exchange couplings $J$ for the copper complex, computed with the broken-spin UHF and GW,
 are close to the experimental results and EOM-SF-CCSD. 
The absolute value of $|J|$ decreases from UHF to GW possibly indicating that capturing of the weak correlation in [Cu$_2$Cl$_6$]$^{2-}$ 
leads to a decrease of $|J|$. 
It explains why GW slightly overestimates $|J|$ with respect to a 
higher-level EOM-SF-CCSD calculation and the available experimental data. 
As discussed in the Ref.\cite{Morokuma:DMRG:biquad_exc:2014}, 
the treatment of weak correlation is important for a correct estimate of the exchange coupling in the iron complexes 
(explaining why UHF and CASSCF numbers are significantly different than the experimental estimate).  
The BS-GW estimate lies in between CASSCF and DMRG estimates. 
The increase in $|J|$ from UHF to GW is consistent with the previous theoretical calculations. 
The significant role of weak correlation is not unique to this iron example 
and has been observed before~\cite{CuAc:Neese:2011,Pokhilko:spinchain}.

The changes in electronic structure can be understood by means of local correlators. 
Davidson and Clark~\cite{Davidson:local_spin:2001} noted that same-center correlators often give unrealistic values 
for the local spin due to closed-shell doubly occupied orbitals. 
The problem is exacerbated when core orbitals are not frozen or when effective core potentials are not used. 
They proposed to use differences of the correlators for the analysis of the energy differences. 
Since GW is not an ensemble respresentable method, 
its total $\braket{S^2}$ and $\braket{N^2}$ also do not have to be realistic. 
However, we see that this error cancels out if the differences are considered. 
For example, squared fluctuation of the number of particles $\braket{(\delta N)^2}$, 
computed for Cu$_2$Cl$_6^{2-}$ with triplet and broken-spin GW solutions are 
$23.0019$ and $23.0046$ respectively. 
While these fluctuations do not describe a realistic system in the low-temperature limit, 
their difference is close to zero, justifying its use in the interpretation of energy differences. 
The difference between $\braket{S^2}$ of the high-spin and broken-spin GW solutions, 
$\braket{S^2}(\text{high spin}) - \braket{S^2}(\text{bs})$, is $0.9995$ and $25.019$ for copper and iron complexes, 
respectively,
which are close to the expected values from the broken-spin physics. 

Tables~S1--S8 in SI show spin and charge correlators computed for 
[Cu$_2$Cl$_6$]$^{2-}$ and [Fe$_2$OCl$_6$]$^{2-}$. 
For [Cu$_2$Cl$_6$]$^{2-}$, the differences between UHF spin correlators are small, 
except the correlator between two different copper centers. 
The difference in spin correlators between copper and neighboring chlorine atoms are about 40 times larger than 
the difference in correlators between copper and distant chlorine atoms. 
The difference in charge correlators is small for all atom pairs indicating that the charge fluctuations for the both electronic states are similar and
confirming the validity of the Heisenberg form of the effective Hamiltonian. 
The inclusion of electronic correlation through GW slightly increases the difference in spin correlators 
between different copper and chlorine centers preserving the Heisenberg form of effective interactions. 
This results only in a small change of the effective exchange coupling. 

UHF difference in spin correlators for [Fe$_2$OCl$_6$]$^{2-}$ is similar to the copper complex.  
While the difference in spin correlators for the same iron centers is small, 
the difference in spin correlators for the adjacent Fe and Cl or O centers is significant. 
This can be rationalized through a larger value of the local spin on an iron center (each Fe has 5 unpaired electrons) resulting in an 
increase of the correlators between iron and chlorine centers. 
The difference in charge correlators is small for all atoms except the oxygen center, 
which means that charge fluctuations on oxygen may influence the value of the effective exchange coupling. 
GW gives a qualitatively different picture. 
First, the difference in the same-center iron spin correlators increases by two orders of magnitude 
and becomes non-negligible ($\Delta \braket{\vec{S}_A \vec{S}_B} = 0.16$). 
Second, the charge correlator between the same and different iron centers increases by an order of magnitude and 
becomes comparable to the one on the oxygen center. 
Consequently, this pronounced effect in charge correlators indicates that the electronic structure of the 
high-spin and low-spin electronic states influences additional configurations than just the main ones, 
resulting in a larger deviation between the broken-spin UHF and GW values. 
Another consequence is the expected deviation from the Heisenberg interaction form. 
Partial differential charge resonance between iron and bridging oxygen centers indicates that 
the interaction between the leading configurations and the charge-transfer configurations is significant, 
which can be wrapped into an effective biquadratic spin interaction 
(as shown through analysis of DFT calculations in  Ref.~\cite{Malrieu:DFT:biquad:2008}). 
This observation explains previous multireference results~\cite{Morokuma:DMRG:biquad_exc:2014}, 
parametrizing the biquadratic interaction. 

To conclude, we applied a local analysis of electronic structure to Green's function methods by 
means of local spin and charge correlators. 
These correlators quantify the covalent and ionic nature of the  Green's function solutions obtained. 
We found that weakly correlated self-consistent GW and GF2 are prone to symmetry breaking in open-shell cases, 
leading to solutions that do not change significantly with temperature. 
The analysis of these solutions allowed us to construct effective Heisenberg Hamiltonians, 
four-spin-four-center interactions, and explain trends in selected single-molecule magnets.

\section*{Acknowledgments}
P.P. and D.Z. acknowledge support from the U.S. Department of Energy under Award No. DE-SC0019374. \\

\section*{Supplementary Material}
FCI temperature dependence of energy and entropy for H$_2$, 
two-particle charge and spin correlators for transition metal complexes, 
Cartesian geometries.

\renewcommand{\baselinestretch}{1.5}

\clearpage
\bibliographystyle{prf}

\end{document}